\documentclass[aps,a4paper,showpacs,prb,twocolumn,superscriptaddress]{revtex4-1}

\usepackage{amsfonts}
\usepackage{amsmath}
\usepackage{amssymb}
\usepackage{graphicx}
\usepackage{verbatim}
\usepackage{color}

\begin{document}

\title{Interaction-induced enhancement of $g$-factor in graphene}

\author{A. V. Volkov}
\email{on\_ton@mail.ru}
\affiliation{Solid State Electronics, ITN, Link\"{o}ping University, 601 74,
Norrk\"{o}ping, Sweden}
\affiliation{Nizhny Novgorod State University,
Gagarin Avenue 23, 603950 Nizhny Novgorod, Russia}

\author{A. A. Shylau}
\email{artsem.shylau@itn.liu.se}\affiliation{Solid State Electronics, ITN, Link\"{o}ping University, 601 74,
Norrk\"{o}ping, Sweden}

\author{I. V. Zozoulenko}
\email{igor.zozoulenko@liu.se}\affiliation{Solid State Electronics, ITN, Link\"{o}ping University, 601 74,
Norrk\"{o}ping, Sweden}

\begin{abstract}
We study the effect of electron interaction on the spin-splitting and the $g$%
-factor in graphene in perpendicular magnetic field using the Hartree and
Hubbard approximations within the Thomas-Fermi model. We found that the $g$%
-factor is enhanced in comparison to its free electron value $g=2$ and
oscillates as a function of the filling factor $\nu $ in the range $2\leq
g^{\ast }\lesssim 4$ reaching maxima at even $\nu $ and minima at odd $\nu $%
. We outline the role of charged impurities in the substrate, which are
shown to suppress the oscillations of the $g^{\ast }$-factor. This effect
becomes especially pronounced with the increase of the impurity
concentration, when the effective $g$-factor becomes independent of the
filling factor reaching a value of $g^{\ast }\approx 2.3$. A relation to the
recent experiment is discussed.
\end{abstract}

\date{\today}
\maketitle

\section{Introduction}

Graphene being subjected to a perpendicular magnetic field exhibits the
unusual quantization of the energy spectrum, which is manifested in a
non-equally spaced sequence of the Landau levels\cite{Goerbig}. In contrast
to conventional two-dimensional electron gas (2DEG) systems, the energy
difference between the lowest Landau levels is large enough allowing
observation of the quantum Hall plateau even at room temperatures\cite%
{Novoselov2007}. Another interesting peculiarity of graphene is the
existence of the 0'th Landau level located precisely at the Dirac point and
equally shared by electrons and holes\cite{Goerbig}. If the magnetic field
is high enough, in addition to the Landau level quantization, the level
splitting due to the Zeeman effect takes place. This kind of splitting was
clearly observed in the recent experiments even for states lying relatively
far from the Dirac point, at the filling factors $\nu =\pm 4.$\cite{Zhao}
The Zeeman splitting is by its nature a one-electron effect, which tells
that a particle possessing a spin degree of freedom acquires the additional
energy in the magnetic field $B$,
\begin{equation}
V_{Z}^{\sigma }=\sigma g\mu _{B}B,  \label{Z}
\end{equation}%
where $\sigma =\pm \frac{1}{2}$ describes two opposite spin states $\uparrow
,\downarrow $; $\mu _{B}$ is the Bohr magneton, $g$ is the free electron
Lande factor ($g$-factor); $g=2$ for graphene. However, experimentally
observed splitting of the Landau levels can not be solely
attributed to the Zeeman effect, as this splitting can also be enhanced by
electron-electron interaction\cite{Ando1974}. The electron-electron
interaction in graphene is especially important at high magnetic fields near
$\nu =0$ when a new insulating state emerges\cite{Zhang2010}. Even though
the nature of this state is still under debate, it is commonly believed that
it is related to the electron-electron interaction\cite{Zhao}.

The enhancement of the spin-splitting due the electron-electron interaction
can be described by introducing a phenomenological effective $g$-factor, $%
g^{\ast },$ which effectively incorporates the interaction effects within
the one-electron description. Calculation of the effective $g$-factor was
originally done for conventional 2DEG systems based on Si MOS\cite{Ando1974}
and GaAs/AlGaAs\cite{Englert1982} structures. It was show that the $g$%
-factor can be enhanced by the electron-electron interaction up to one order
of magnitude in comparison to its bare value\cite{Englert1982} and
oscillates as a function of a carrier density\cite{Ando1974,Englert1982}.
Interaction induced spin-splitting was extensively studied in confined 2DEG
structures such as quantum wires \cite%
{Kinaret,Dempsey,Tokura,takis2002,Stoof,Ihnatsenka_wire1,Ihnatsenka_wire_comp_strips,IhnatsenkaCEOQW,IhnatsenkaMarcus}%
. It was also argued that interaction-induced spontaneous spin-splitting can
take place in 2DEG systems even in the absence of magnetic field\cite%
{Ghosh,Goni,Evaldsson}.

The enhancement of the effective $g$-factor was also observed in carbon
based systems. In graphite the effective $g$-factor is reported to be $%
g^{\ast }\approx 2.5.$\cite{Schneider} Recently, Kurganova \textit{et. al.}%
\cite{kurganova} performed measurements of the effective $g$-factor in
graphene. It was found to be $g^{\ast }=2.7\pm 0.2$, which is larger than
its non-interacting value $g=2$. This indicates that electron-electron
interaction effects play an important role and should be taken into account
for explanation of the enhanced spin-splitting. Motivated by this experiment
we use the Thomas-Fermi approach to study the spin-splitting in realistic
two-dimensional graphene sheets in perpendicular magnetic field situated on
a dielectric surface and subjected to a smooth confining potential due
to charged impurities. (Note that the enhancement of effective $g$-factor in
ideal graphene nanoribbons has been recently studied by Ihnatsenka \textit{%
et. al}\cite{Ihnatsenka2012}). The paper is organized as follows. Sec. II
presents the model, where we specify the system at hand and define the
Hamiltonian. In Sec. III we discussed the obtained results and provide an
explanation for the observed behavior of $g^{\ast }$-factor. Sec. IV
contains the conclusions.

\section{Model}

\begin{figure}[tbh]
\includegraphics[keepaspectratio,width= \columnwidth]{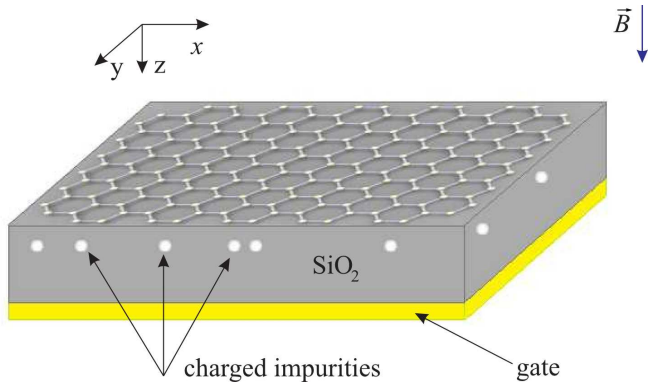}
\caption{(Color online) Schematic illustration of the studied structure. A
graphene sheet is located on an insulating substrate of the width $d$
separating it from a metallic gate. The substrate is contaminated by charged
impurities with $q=\pm 1$ situated at the distanace $h=1$ nm apart from the graphene layer.}
\label{fig:device}
\end{figure}
We consider a system depicted in Fig. \ref{fig:device}, consisting of a
graphene sheet located on an insulating substrate of the width $d$ with the
dielectric constant $\epsilon _{r}.$ (We choose $\epsilon _{r}=3.9$
corresponding to SiO$_{2}$). A metallic back gate is used to tune the
carrier density by varying the gate voltage $V_{g}$. We assume the charged
impurities with the concentration $n_{i}$ are randomly distributed in the
substrate at the distance $h=1$ nm apart from the graphene layer\cite%
{dassarma}. The whole system is subjected to the perpendicular magnetic
field $B$. In order to find the ground-state carrier density, we use the
Thomas-Fermi approximation with the local relation\cite%
{Gerthards1994,Gerthards1997,Hannes}
\begin{equation}
n_{\sigma }(\mathbf{r})=\left\{
\begin{array}{ll}
\int\limits_{V^{\sigma }(\mathbf{r})}^{\infty }\rho ^{\sigma }(E-V^{\sigma }(%
\mathbf{r}))f_{FD}^{e}(E-E_{F})dE & \text{(electrons)} \\
&  \\
\int\limits_{-\infty }^{V^{\sigma }(\mathbf{r})}\rho ^{\sigma }(E-V^{\sigma
}(\mathbf{r}))f_{FD}^{h}(E-E_{F})dE & \text{(holes)} \\
&
\end{array}%
\right.   \label{n}
\end{equation}%
between the spin-dependent carrier density $n_{\sigma }(\mathbf{r})$ of the
graphene and the total potential energy $V^{\sigma }(\mathbf{r})$. Here $%
f_{FD}^{e}(E-E_{F})=1/\left( \exp \left( \frac{E-E_{F}}{k_{B}T}\right)
+1\right) $ and $f_{FD}^{h}(E,\mu )=1-$ $f_{FD}^{e}(E-E_{F})$ are the
Fermi-Dirac distribution functions for electrons and holes respectively, $%
E_{F}=eV_{g}$ is the Fermi energy. The Landau density of state in graphene
is given by\cite{Goerbig}
\begin{equation}
\rho ^{\sigma }(E)=\left\{
\begin{array}{ll}
\sum\limits_{i=0}^{\infty }\frac{g_{\nu }}{2\pi l_{B}^{2}}\delta (E-\hbar
\omega _{c}\sqrt{i}) & \text{(electrons)} \\
&  \\
\sum\limits_{i=0}^{\infty }\frac{g_{\nu }}{2\pi l_{B}^{2}}\delta (E+\hbar
\omega _{c}\sqrt{i}) & \text{(holes),} \\
&
\end{array}%
\right.   \label{rho}
\end{equation}%
where $\omega _{c}=\sqrt{2}v_{F}/l_{B}$ is the cyclotron frequency, $l_{B}=%
\sqrt{\hbar /eB}$ is the magnetic length, $v_{F}$ is the Fermi velocity in
graphene; the factor $g_{v}=2$ takes into account the valley degeneracy for
all levels except of the zeroth one. The zeroth Landau level belongs both to
electrons and holes which we take into account by setting $g_{v}=1$.
According to Eq. (\ref{n}), the carrier density $n_{\sigma }(\mathbf{r})$ at
the position $\mathbf{r}$ depends on the total potential only at that
position.

The total potential
\begin{equation}
V^{\sigma }(\mathbf{r})=V_{H}(\mathbf{r})+V_{U}^{\sigma }(\mathbf{r}%
)+V_{Z}^{\sigma }+V_{imp}(\mathbf{r})  \label{V_tot}
\end{equation}%
is a sum of the Hartree, Hubbard, Zeeman and the external potential produced
by the impurities. The Hartree potential is given by\cite{Shylau_Cap,FR},
\begin{equation}
V_{H}(\mathbf{r})=\frac{e^{2}}{4\pi \epsilon _{0}\epsilon _{r}}\sum_{\mathbf{%
r}^{^{\prime }}\neq \mathbf{r}}n(\mathbf{r}^{\prime })\left( \frac{1}{|%
\mathbf{r}-\mathbf{r}^{\prime }|}-\frac{1}{\sqrt{|\mathbf{r}-\mathbf{r}%
^{\prime }|^{2}+4d^{2}}}\right) ,  \label{V_H}
\end{equation}%
where $n(\mathbf{r})=n_{\uparrow }(\mathbf{r})+n_{\downarrow }(\mathbf{r})$
is the local carrier density, and the second term describes a contribution
from the mirror charges\cite{Hartree}. The second term in Eq. (\ref{V_tot})
is the standard Hubbard potential which is shown to describe carbon electron
systems in a good agreement with the first-principles calculations\cite%
{FR,Wehling}
\begin{equation}
V_{U}^{\sigma }(\mathbf{r})=Un^{\sigma ^{\prime }}(\mathbf{r})S_{a},
\label{Hubbard}
\end{equation}%
where $U$ is the effective Hubbard constant and $S_{a}=\frac{3\sqrt{3}}{4}%
a^{2}$ is the area of unit cell of graphene($a\approx 0.142$ nm is the
carbon-carbon distance). In our work we use $U=9.3$ eV which has been
recently calculated within the constrained random phase approximation\cite%
{Wehling}. The third term Eq. (\ref{V_tot}) is the Zeeman energy given by
Eq. (\ref{Z}). The last term in Eq. (\ref{V_tot}) corresponds to the
potential due to charged impurities and is given by
\begin{eqnarray}
&&V_{imp}(\mathbf{r})=\frac{e^{2}}{4\pi \epsilon _{0}\epsilon _{r}}\times
\notag \\
&&\times \sum_{i=1}^{N_{imp}}\left( \frac{q_{i}}{|\mathbf{r}-\mathbf{r}%
_{i}|^{2}+h^{2}}-\frac{q_{i}}{\sqrt{|\mathbf{r}-\mathbf{r}%
_{i}|^{2}+(2d-h)^{2}}}\right) ,  \notag \\
&&
\end{eqnarray}%
where the summation is performed over charged impurities in the dielectric; $%
\mathbf{r}_{i}$ is the coordinate in the graphene plane of the projection of
the $i$-th impurity of the charge $q_{i}$ situated at the distance $h$ from
the plane. Equations (\ref{n}) and (\ref{V_tot}) are solved
self-consistently until a convergence is achieved.

We define the effective $g$-factor as follows,
\begin{equation}
g^{\ast }\mu _{B}B=\left\langle V^{\uparrow }(\mathbf{r})-V^{\downarrow }(%
\mathbf{r})\right\rangle ,  \label{g_def}
\end{equation}%
which assumes that spin-splitting in the system is caused by the Zeeman
term, Eq. (\ref{Z}), where the free electron value $g$ is replaced by the
effective $g$-factor, $g^{\ast }.$ (If the Hubbard interaction is absent, $%
U=0$, then apparently $g^{\ast }=g)$. Substituting Eq.(\ref{V_tot}) into Eq.(%
\ref{g_def}), we arrived at the equation used to calculate $g^{\ast },$
\begin{equation}
g^{\ast }=g+\frac{U}{\mu _{B}B}\left\langle n_{\downarrow }(\mathbf{r}%
)-n_{\uparrow }(\mathbf{r})\right\rangle ,  \label{g_eff}
\end{equation}%
where $\langle \ldots \rangle $ denotes spatial averaging over the graphene
lattice sites.

\section{Results and discussion}

\begin{figure}[tbh]
\includegraphics[keepaspectratio,width=\columnwidth]{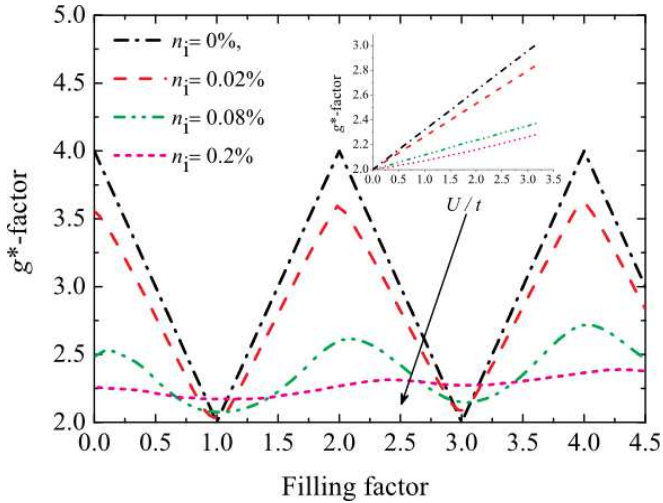}
\caption{(Color online) The effective $g$-factor as a function of the
filling factor $\protect\nu $ for different concentrations of charged
impurities, $n_{i}=0\%,0.02\%,0.08\%,0.2\%$, at the constant perpendicular
magnetic field $B=35$T. Inset: the dependence of $g^{\ast }$ on the Hubbard
constant $U$ for the fixed $\protect\nu =2.5$. All the calculations are done
at the temperature $T=4$ K.}
\label{fig:g}
\end{figure}

Figure \ref{fig:g} presents the central result of the paper. It shows the
effective $g$-factor as a function of the filling factor $\nu =\frac{n}{n_{B}}$ ($n_{B}=\frac{1}{\pi l_{B}^{2}}$, note that our definition of $n_B$ includes a factor of 2 accounting for the valley degeneracy) calculated for different concentrations
of impurities.
The dependence $g^{\ast }=g^{\ast }(\nu )$ exhibits two main features.
First, the effective $g$-factor is enhanced ($g^{\ast }>g$) and oscillates
in the range $2\leq g^{\ast }\lesssim 4$ achieving its maximal values at
even filling factors $\nu =0,2,4,\ldots $, while having the minima at odd
filling factors $\nu =1,3,\ldots $. Second, the increase of the impurity
concentration suppresses the enhancement as well as the oscillatory behavior
of $g^{\ast }$, such that for high $n_{i}$ the effective $g$-factor becomes
only weakly dependent on $\nu $ reaching the value $g^{\ast }\approx 2.3.$

\begin{figure}[tbh]
\includegraphics[keepaspectratio,width=\columnwidth]{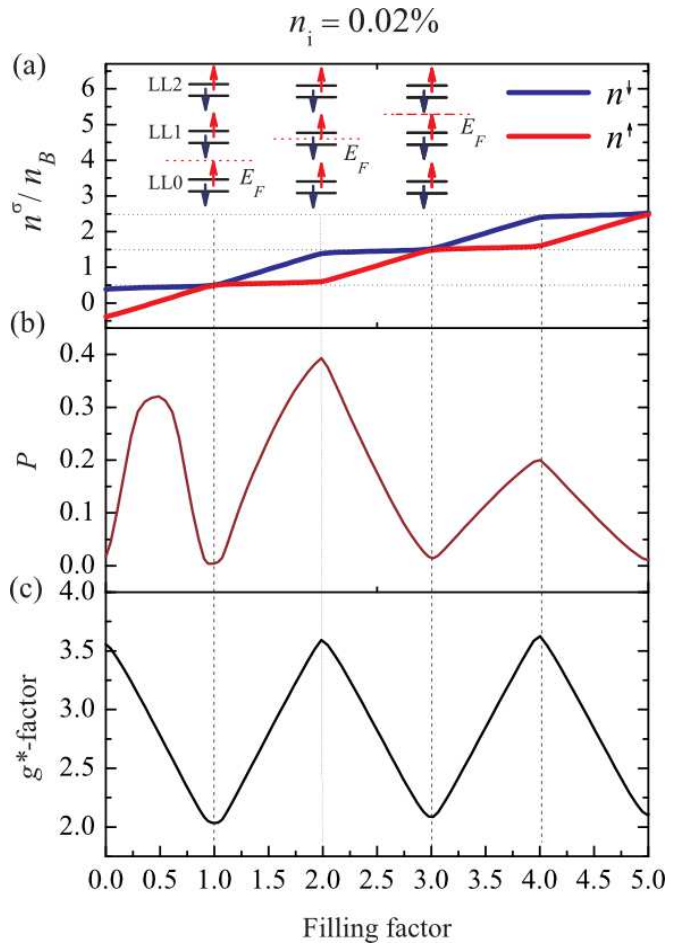}
\caption{(Color online) The dependence of (a) the charge concentration, (b)
the polarization and (c) the effective $g$-factor on the filling factor in
almost ideal system (i.e. in a graphene sheet with the low concentration of
charged impurities in the substrate, $n_{i}=0.02\%$).}
\label{fig:g_0.02}
\end{figure}

In order to understand the observed behavior let us first consider in
details the case of a low concentration of impurities shown on Fig. \ref%
{fig:g_0.02}, where $g^{\ast }$-factor dependence for $n_{i}=0.02\%$ is
complemented by the spin-density and the polarization dependencies. For
small $n_{i}$ the effect of impurities is small and the filling factor can
be directly related to the number of the occupied Landau levels of an ideal
system (i.e. without impurities). For a fixed value of the magnetic field
the increase of the filling factor corresponds to the increase of charge
density through subsequent population of the Landau levels. As seen in Fig. %
\ref{fig:g_0.02}, the total spin polarization $P=\left\langle P(\mathbf{r}%
)\right\rangle =\left\langle \frac{|n_{\downarrow }(\mathbf{r}%
)|-|n_{\uparrow }(\mathbf{r})|}{|n_{\downarrow }(\mathbf{r})|+|n_{\uparrow }(%
\mathbf{r})|}\right\rangle $ exhibits the same qualitative behavior as the $%
g $-factor (except for $\nu =0$, which will be discussed below). At $\nu =1$
the $g^{\ast }$-factor reaches its minimal value $g^{\ast }=g$. In this case
the Fermi energy is located in between the 0'th Landau level (LL0) and the
1'st Landau level (LL1), i.e. the LL0 is fully occupied, while LL1 is
completely empty (see inset in Fig. \ref{fig:g_0.02}). This gives rise to
the equal spin-up and spin-down densities and hence to the zero spin
polarization. The increase of the filling factor in the range $1<\nu <2$
leads to gradual population of 1'st spin-down ($\downarrow $) Landau level
(LL1($\downarrow $)) and, in turn, to the increase of $n_{\downarrow }$,
while $n_{\uparrow }$ does not change. (Note that even though in our model
the DOS is given by the delta functions, it is effectively smeared out by a
non-zero temperature, which results in a smooth change of the charge
densities.) Since the difference $n_{\downarrow }-n_{\uparrow }$ increases,
according to Eq. (\ref{g_eff}) $g^{\ast }$ grows and reaches its maximum $%
g^{\ast }\approx 3.5$ at $\nu =2$, when the Fermi energy lies in the middle
of two spin-split levels corresponding to the same Landau level (LL1). The
enhancement of the effective $g$-factor in comparison to its noninteracting
value is apparently caused by the Hubbard term in Eq. (\ref{V_tot}). The
Hubbard interaction enhances the spin-splitting triggered by the Zeeman
interaction giving rise to $g^{\ast }>g$.

When the filling factor is further increased from $\nu =2$ to $\nu =3$, i.e.
the Fermi energy is shifted towards higher energies, the population of the
spin-up ($\uparrow $) level belonging to LL1 gradually grows, while the
density of the spin-down electrons ($\downarrow $) belonging to the same LL1
remains unchanged as the later level remains completely filled. Eventually,
at $\nu =3$ the spin densities become equal, $n_{\downarrow }\approx
n_{\uparrow }$, the system is not spin-polarized ($P=0$) and the effective $%
g $-factor again reaches its minimum $g^{\ast }=g$. The same physics is
responsible for similar oscillatory behavior of the effective $g$-factor and
the polarization for higher filling factors.
\begin{figure}[tbh]
\includegraphics[keepaspectratio,width=\columnwidth]{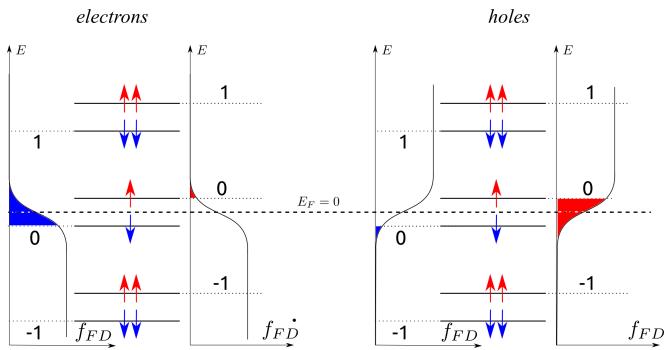}
\caption{(Color online) Schematic illustration of Landau levels population
at $\protect\nu =0$ for electrons and holes. Shaded regions (red and blue)
correspond to states occupied in the LL0 by spin-up and spin-down electrons.}
\label{fig:LL0}
\end{figure}

The dependencies of the effective $g$-factor and the polarization are
qualitatively different for $\nu <1$. Namely, the polarization drops to zero
at $\nu =0$, while the effective $g$-factor reaches its maximum, see Fig. %
\ref{fig:g_0.02}(b),(c). This is in contrast to all other even filling
factors when both $g^{\ast }$ and $P$ exhibit maxima. This can be understood
as follows. In contrast to other Landau levels, LL0 is equally shared by
electrons and holes at $E_{F}=0$, which is a distinct feature of graphene.
As illustrated in Fig. \ref{fig:LL0}, when the magnetic field is high
enough, i.e. the spin-split levels are well resolved, electrons
predominantly populate the LL0($\downarrow $) state, while LL0($\uparrow $)
is mostly occupied by holes. As a result, $n_{\downarrow }=-n_{\uparrow }$,
and therefore the effective $g$-factor reaches the maximum because of the
Hubbard term $\sim U(n_{\downarrow }-n_{\uparrow })=2Un_{\downarrow }$. On
the other hand, at $\nu =0$ the graphene is electrically neutral, $%
n=n_{\downarrow }+n_{\uparrow }=0$, and spin-polarization is absent, $P=0,$
since $|n_{\downarrow }|=|n_{\uparrow }|$. Note that the effect of
electron-electron interaction on spin splitting in graphene nanoribbons at $%
\nu \approx 0$ was discussed in Ref.[\cite{Shylau2011}].
\begin{figure}[tbh]
\includegraphics[keepaspectratio,width=\columnwidth]{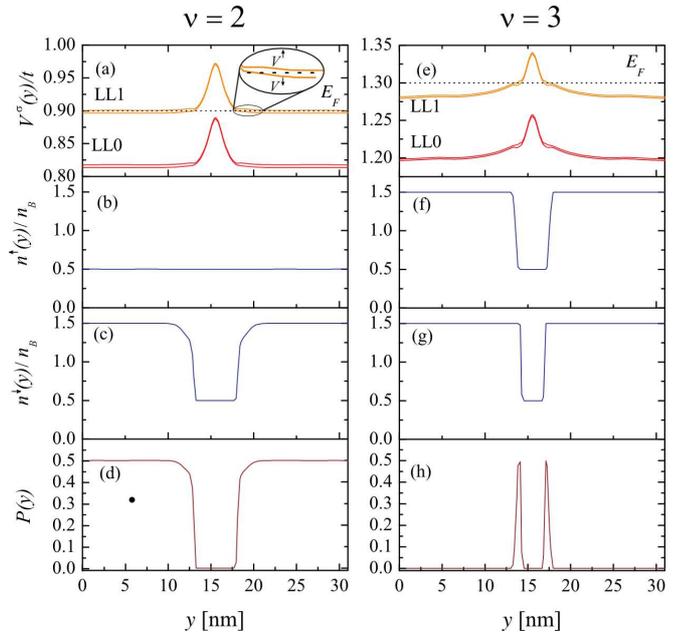}
\caption{(Color online) Distribution of (a),(e) the self-consistent
potential, (b),(f) spin-up and (c),(g) spin-down electron density, (d),(h)
the spin polarization for a single impurity at different filing factors $%
\protect\nu =2$ and $\protect\nu =3$ (left and right columns respectively). }
\label{fig:1imp}
\end{figure}

The above analysis is strictly speaking applicable only for ideal graphene,
when $n_{i}=0$. In this case, the range of $g$-factor oscillations can be
easily estimated from Eq. (\ref{g_eff}). At odd filling factors, when $%
n_{\downarrow }=n_{\uparrow }$, Eq. (\ref{g_eff}) gives $g_{\text{min}%
}^{\ast }=g=2$. At even filling factors, when $E_{F}$ lies between two
spin-split levels of a given Landau level, for the chosen parameters $U$ and
$B,$ the effective $g$-factor $g_{\text{max}}^{\ast }=2+\frac{US_{a}}{\pi
l_{B}^{2}\mu _{B}}\approx 4$, which is in accordance with our numerical
calculations (Fig. \ref{fig:g}, $n_{i}=0$). However, in the presence of
impurities, this is not the case anymore, as the oscillations of the
effective $g$-factor get suppressed and $g^{\ast }$ never reaches $g_{\text{%
max}}^{\ast }$ and always stays larger than $g_{\text{min}}^{\ast }$, see
Fig. \ref{fig:g}.

In order to explain the influence of impurities on the $g$-factor, let us
now consider a system consisting of a single repulsive impurity only. Figure %
\ref{fig:1imp} (a) shows the cross section of the self-consistent potentials
$V^{\uparrow }$ and $V^{\downarrow }$ for spin-up and spin-down electrons
respectively. The LL0($\downarrow ,\uparrow $) coincides with the
self-consistent potential $V^{\downarrow ,\uparrow }$, while the positions
of the LL1($\downarrow ,\uparrow $) are given by $V^{\downarrow ,\uparrow
}+\hbar \omega _{c}$. We have chosen two representative values of the
filling factor, namely, $\nu =2$ and $\nu =3$ corresponding to maximum and
minimum values of the effective $g$-factor.

At $\nu =2$, which in ideal graphene corresponds to the almost occupied
spin-down and almost empty spin-up states of the LL1, $g^{\ast }$ reaches
the maximal value. Figure \ref{fig:1imp} shows that the LL1($\downarrow $)
is pinned to the Fermi energy $E_{F}$. (For the effect of pinning of $E_{F}$
within the Landau levels see e.g. Ref. \cite{Davies}). The states lying in
the interval $|E-E_{F}|<2\pi k_{B}T$ are partially filed $0<f_{FD}<1$ and
therefore the electron density can be redistributed under an influence of an
external potential. These states represents the compressible strips\cite%
{Chklovskii1992}, which in our case extends over the whole system (except of
the impurity region). The presence of negative impurity leads to the
distortion of the potential as depicted in Fig. \ref{fig:1imp}(a). As a
result, in the impurity region the LL1($\downarrow $) raises above $E_{F}$
and this state becomes depopulated, Fig. \ref{fig:1imp}(c). (Note that LL1($%
\uparrow $) is practically depopulated even in an absence of the impurity,
Fig. \ref{fig:1imp}(b)). As a result, the spin density difference, $%
n_{\downarrow }-n_{\uparrow },$ decreases in the impurity region, which
apparently leads to the decrease of $P$ and $g^{\ast }$ in comparison to
ideal graphene, see Fig. \ref{fig:1imp}(d).

On the other hand, the influence of the impurity is opposite for odd $\nu $.
At $\nu =3$ the system is predominantly in a unpolarized state, which is
manifested by the minimum of $g^{\ast }$. However, the distortion of the
potential due to the impurity gives rise to the formation of a compressible
strip around the impurity, where $E_{F}$ intersects the LL1. This is clearly
seen in Fig. \ref{fig:1imp} (e) where the compressible strip corresponds to
regions where the potential is flat because of the pinning to $E_{F}$ within
the energy window $|E-E_{F}|<2\pi k_{B}T$ (where $0<f_{FD}<1).$ Because of
the partial filling of the compressible strip, the electron density there
can be easily redistributed there. As a result, the Hubbard interaction
pushes up and depopulates the LL1($\uparrow $) while the LL1($\downarrow $)
remains populated, see Figs. \ref{fig:1imp}(f),(g). This leads to a local
spin polarization around the impurity as illustrated in Fig.(\ref{fig:1imp}%
)(h). therefore, the overall polarization is no longer zero, $\langle P(%
\mathbf{r})\rangle >0$ and, hence, the effective $g$-factor does not drop to
the minimum value, remaining $g^{\ast }>g_{\text{min}}^{\ast }$.

Summarizing, the influence of a single impurity is twofold: when the system
is predominantly spin-polarized (even $\nu $), the impurity decreases the
average polarization and the effective $g$-factor by locally pushing-up the
Landau levels and depopulating them; in the opposite case of a predominantly
non-polarized system (odd $\nu $), the impurity leads to the local
formation of the spin-polarized compressible strips, which instead increases
the average polarization and the effective $g$-factor.

\begin{figure*}[tbh]
\includegraphics[keepaspectratio,width=1.5\columnwidth]{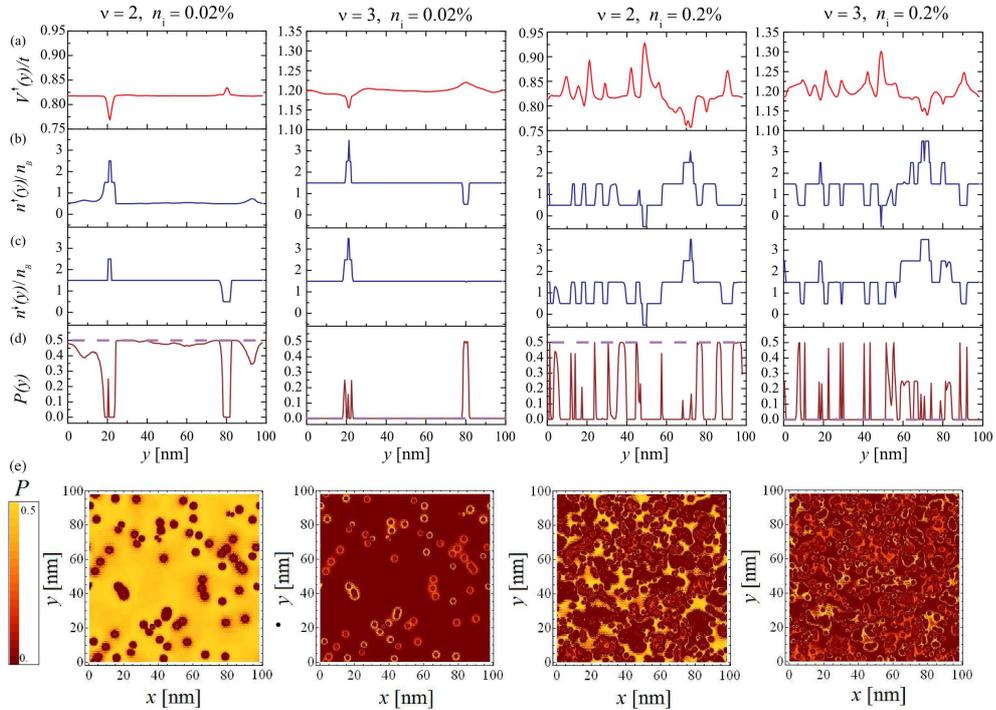}
\caption{(Color online). The spin resolved potential, densities and
polarization for different concentration charged of impurities($%
n_{i}=0.2\%,0.02\%$) and for different filling factors ($\protect\nu =2,3$).
The one-dimensional plots of (a) $V^{\uparrow }(y),$ (b) the spin-up and (c)
spin-down charge densities, and (d) the spin polarization $P(y)$ as a
functions of $y$ for $x=50$ nm. Dashed lines correspond to the ideal system
(without impurities). (e) The 2D plot of the spatially resolved spin
polarization $P(x,y)$ in a graphene sheet. The system parameters are $%
N_{x}=800$, $N_{y}=461$, $d=10$ nm, $B=50$ T.}
\label{fig:SpinPolarization}
\end{figure*}
Having understood the effect of a single impurity on the average
polarization and the effective $g$-factor it is straightforward to
generalize the obtained results for an arbitrary concentration of
impurities. The higher concentration $n_{i}$, the larger influence of
impurities on the average value of the spin-polarization and the effective $g$-factor. As a
result, an increase of the impurity concentration leads to the suppression
of the amplitude of oscillations as shown in Fig. \ref{fig:g}.

Note that for a sufficiently large impurity concentration (in our case $%
n_{i}=0.2\%$), the oscillations of $g^{\ast }$ get practically suppressed
and $g^{\ast }$ becomes rather independent on the filling factor, see Fig. %
\ref{fig:g}. This effect can be understood from a comparison of two distinct
cases of low and high impurity concentration, $n_{i}=0.02\%$ and $%
n_{i}=0.2\% $, see Fig. \ref{fig:SpinPolarization}. When the impurity
concentration is low ($n_{i}=0.02\%$, two left columns in Fig. \ref%
{fig:SpinPolarization}), the self-consistent potentials produced by
different impurities do not overlap and the system can be treated as an
assembly of independent impurities. (The potential is flat everywhere
besides narrow regions close to the impurities, see (a)-panels for $%
n_{i}=0.02\%$ in Fig. \ref{fig:SpinPolarization}). At $\nu =2$ the presence
of impurities decreases locally polarization (dips on (c)-panel), while at $%
\nu =3$ the local polarization increases (peaks on (c)-panel).

However, when the impurity concentration is high ($n_{i}=0.2\%$, two right
columns in Fig. \ref{fig:SpinPolarization}), the potentials produced by
different impurities start to overlap and the analysis in terms of a single
impurity is no longer justified. A given value of the filling factor can not
be associated with a certain number of the Landau levels, since the
potential is strongly distorted in comparison to the ideal case ((a)-panels
for $n_{i}=0.2\%$ in Fig. \ref{fig:SpinPolarization}) and therefore
electrons occupy different Landau levels ((b) and (c)-panels for $%
n_{i}=0.2\% $ in Fig. \ref{fig:SpinPolarization}). In fact, the deviations
in the potential and densities from those of the ideal case become so
significant, so the difference between the cases of $\nu =2$ and $\nu =3$ is
practically washed out (c.f. two right columns in Fig. \ref%
{fig:SpinPolarization}). As a result, the average value of the polarization
and the effective $g$-factor becomes practically independent of the filling
factor.

In the model used in our calculation the enhancement of the $g$-factor is
caused by the Hubbard term in the potential, Eqs. (\ref{Hubbard}) and (\ref%
{g_eff}). Let us briefly discuss how the calculated value of $g^{\ast }$
depends on the Hubbard constant $U$. While we used value $U\approx 3.5t$\cite%
{Wehling}, the current literature reports various estimations of $U$ in the
range $0.5t\lesssim U\lesssim 2t,$\cite{Yazyev, Jung,Tao} where $t\approx
2.7 $ eV is the hopping integral in the standard $p$-orbital tight-binding
Hamiltonian\cite{Goerbig}. We calculated the dependencies $g^{\ast }=g^{\ast
}(\nu )$ for different values of the parameter $U$ and found that the
results show the same qualitative behaviour and the calculated value of $%
g^{\ast }$ scales linearly with $U.$ This is illustrated in the inset to
Fig. \ref{fig:g} which shows a dependence of the effective $g$-factor on the
Hubbard constant for a representative value of $\nu =2.5.$

Let us now discuss the relation of our findings to the recent experiment.
Measurements done by Kurganova \textit{et al.}\cite{kurganova} exhibit the
enhancement of the effective spin-splitting leading to the effective $g$%
-factor $g^{\ast }=2.7\pm 0.2$. Also, the enhanced effective $g$-factor was
found to be practically independent on $\nu $. Our calculations show that
for low impurity concentrations, $g^{\ast }$ exhibits a pronounced
oscillatory behavior in the range $2\leq g^{\ast }\lesssim 4,$ and it
becomes rather independent of $\nu $ for larger $n_{i}$ reaching a saturated
value $g^{\ast }\sim 2.3.$ Our calculations therefore strongly suggest that
impurities always present in realistic samples play an essential role in
suppressing the oscillatory behavior of $g^{\ast }.$ Note that in real
systems the oscillations of $g^{\ast }$ can be smoothed by a number of
additional factors. The measurements of Kurganova \textit{et. al }\cite%
{kurganova} were performed in a tilted magnetic fields and at large filling
factors $\nu >6$. In this case the distance between the adjacent Landau
levels is comparable to the Zeeman splitting which results in stronger
overlap of the successive Landau levels and eventually leads to an
additional smearing of $g^{\ast }$. Therefore our calculations motivate for
further studies of the effective $g$-factor close to $\nu =0$, where the
oscillatory behavior of $g^{\ast }$ is expected to be more pronounced. Our
finding also indicate that the oscillatory behavior of the effective $g$%
-factor is expected to be more pronounced in suspended samples where the
influence of charged impurities will be much less important.

Finally, it is noteworthy that spin-splitting of in graphene\cite{Lundeberg}
and graphene quantum dots\cite{Guttinger} was also experimentally studied in
a parallel magnetic filed. It was concluded that in this case the effective $%
g$-factor does differ from its free-electron value. This can be explained by
the fact that in the parallel field the Landau levels do not form and
therefore the interaction induced enhancement of the $g^{\ast }$-factor is
small.

\section{Conclusions}

In this work we employed the Thomas-Fermi approximation in order to study
the effective $g$-factor in graphene in the presence of a perpendicular
magnetic field taking into account the effect of charged impurities in the
substrate. We found that electron-electron interaction leads to the
enhancement of the spin splitting, which is characterized by the increase of
the effective $g$-factor. We showed that for low impurity concentration $%
g^{\ast }$ oscillates as a function of the filling factor $\nu $ in the
range from $g_{\text{min}}^{\ast }=2$ to $g_{\text{max}}^{\ast }\approx 4$
reaching maxima at even filling factors and minima at odd ones. Finally, we
outlined the influence of impurities on the spin-splitting and demonstrated
that the increase of the impurity concentration leads to the suppression of
the oscillation amplitude and to a saturation of the the effective $g$%
-factor around a value of $g^{\ast }\approx 2.3$.

\begin{acknowledgments}
We acknowledge a support of the Swedish 
Research Council (VR) and the Swedish Institute (SI). A.V.V.  also acknowledges  the Dynasty foundation for  financial support. The authors are grateful to V. Gusynin for critical reading of the manuscript.
\end{acknowledgments}


\begin{thebibliography}{99}
\bibitem{Goerbig} M. O. Goerbig, Rev. Mod. Phys. \textbf{83}, 1193 (2011).

\bibitem{Novoselov2007} K. S. Novoselov, Z. Jiang, Y. Zhang, S. V. Morozov,
H. L. Stormer, U. Zeitler, J. C. Maan, G. S. Boebinger, P. Kim, and A. K.
Geim, Science 315, 1379 (2007).

\bibitem{Zhao} Yue Zhao, Paul Cadden-Zimansky, Fereshte Ghahari, and Philip
Kim, Phys. Rev. Lett. \textbf{108}, 106804 (2012).

\bibitem{Ando1974} T. Ando and Y. Uemura, J. Phys. Soc. Jpn. \textbf{37},
1044 (1974).

\bibitem{Zhang2010} L. Zhang, Y. Zhang, M. Khodas, T. Valla, and I. A.
Zaliznyak, Phys. Rev. Lett. \textbf{105}, 046804 (2010).

\bibitem{Englert1982} T. Englert, D. Tsui, A. Gossard, and C. Uihlein, Surf.
Sci. 113, 295 (1982).

\bibitem{Kinaret} J. M. Kinaret and P. A. Lee, Phys. Rev. B \textbf{42},
11768 (1990).

\bibitem{Dempsey} J. Dempsey, B. Y. Gelfand, and B. I. Halperin, Phys. Rev.
Lett. \textbf{70}, 3639 (1993).

\bibitem{Tokura} Y. Tokura and S. Tarucha, Phys. Rev. B \textbf{50}, 10981
(1994).

\bibitem{takis2002} Z. Zhang and P. Vasilopoulos, Phys. Rev. B \textbf{66},
205322 (2002).

\bibitem{Stoof} T. H. Stoof and G. E. W. Bauer, Phys. Rev. B \textbf{52},
12143 (1995).

\bibitem{Ihnatsenka_wire1} S. Ihnatsenka and I. V. Zozoulenko, Phys. Rev. B
\textbf{73}, 075331 (2006).

\bibitem{Ihnatsenka_wire_comp_strips} S. Ihnatsenka and I. V. Zozoulenko,
Phys. Rev. B \textbf{73}, 155314 (2006).

\bibitem{IhnatsenkaCEOQW} S. Ihnatsenka and I. V. Zozoulenko, Phys. Rev. B
\textbf{74}, 075320 (2006).

\bibitem{IhnatsenkaMarcus} S. Ihnatsenka and I. V. Zozoulenko, Phys. Rev. B
\textbf{78}, 035340 (2008).

\bibitem{Ghosh} A. Ghosh, C. J. B. Ford, M. Pepper, H. E. Beere, and D. A.
Ritchie, Phys. Rev. Lett. \textbf{92}, 116601 (2004).

\bibitem{Goni} A. R. Go\~{n}i, P. Giudici, F. A. Reboredo, C. R. Proetto, C.
Thomsen, K. Eberl, and M. Hauser, Phys. Rev. B \textbf{70}, 195331 (2004).

\bibitem{Evaldsson} M. Evaldsson, S. Ihnatsenka, and I. V. Zozoulenko, Phys.
Rev. B \textbf{77}, 165306 (2008).

\bibitem{Schneider} J. M. Schneider, N. A. Goncharuk, P. Vasek, P. Svoboda,
Z. Vyborny, L. Smrcka, M. Orlita, M. Potemski, and D. K. Maude, Phys. Rev. B
\textbf{81}, 195204 (2010).

\bibitem{kurganova} E. V. Kurganova, H. J. van Elferen, A. McCollam, L. A.
Ponomarenko, K. S. Novoselov, A. Veligura, B. J. van Wees, J. C. Maan, and
U. Zeitler, Phys. Rev. B, \textbf{84}, 121407, (2011).

\bibitem{Ihnatsenka2012} S. Ihnatsenka and I. V. Zozoulenko,
arXiv:1206.6251v1 [cond-mat.mes-hall] (2012).

\bibitem{dassarma} Enrico Rossi and S. Das Sarma, Phys. Rev. Lett. \textbf{%
101}, 166803 (2008).

\bibitem{Gerthards1994} K. Lier and R. R. Gerhardts, Phys. Rev. B \textbf{50}%
, 7757 (1994).

\bibitem{Gerthards1997} J. H. Oh and R. R. Gerhardts, Phys. Rev. B \textbf{56%
}, 13519 (1997).

\bibitem{Hannes} W.-R. Hannes, M. Jonson and M. Titov, Phys. Rev. B \textbf{%
84}, 045414 (2011).

\bibitem{Shylau_Cap} A. A. Shylau, J. W. Klos, and I. V. Zozoulenko, Phys.
Rev. B \textbf{80}, 205402 (2009).

\bibitem{FR} J. Fern\'{a}ndez-Rossier, J. J. Palacios, and L. Brey, Phys.
Rev. B \textbf{75}, 205441 (2007).

\bibitem{Hartree} To facilitate calculations of $V(\mathbf{r})$ the
summation in Eq.(\ref{V_H}) was performed numerically in the region $|%
\mathbf{r}-\mathbf{r}^{\prime }|<R$. Outside this region the charge density
was assumed to be uniform $n(\mathbf{r})=\langle n\rangle $, such that that
the additional potential produced by $\langle n\rangle $ by charges in the
this region was evaluated analytically.

\bibitem{Wehling} T. O. Wehling, E. \c{S}a\c{s}io\u{g}lu, C. Friedrich, A.
I. Lichtenstein, M. I. Katsnelson and S. Bl\"{u}gel, Phys. Rev. Lett. 106, 236805 (2011).

\bibitem{Shylau2011} A. A. Shylau and I. V. Zozoulenko, Phys. Rev. B \textbf{%
84}, 075407 (2011).

\bibitem{Davies} J. H. Davies, \textit{The physics of low-dimensional
semiconductors: an introduction}, (Cambridge university press, Cambridge,
1998).

\bibitem{Chklovskii1992} D. B. Chklovskii, B. I. Shklovskii, and L. I.
Glazman, Phys. Rev. B \textbf{46}, 4026 (1992).

\bibitem{Yazyev} O. V. Yazyev, Phys. Rev. Lett. \textbf{101}, 037203 (2008).

\bibitem{Jung} J. Jung and A. H. MacDonald, Phys. Rev. B \textbf{80}, 235417 (2009).

\bibitem{Tao} C. Tao, L. Jiao, O. V. Yazyev, Y.-C. Chen, J. Feng, X. Zhang,
R. B. Capaz, J. M. Tour, A. Zettl, S. G. Louie, H. Dai, and M. F. Crommie,
Nat. Phys. 7, 616 (2011).

\bibitem{Lundeberg} Mark B. Lundeberg and Joshua A. Folk, Nat. Phys. \textbf{5}, 894 (2009)

\bibitem{Guttinger} J. Guttinger, T. Frey, C. Stampfer, T. Ihn, and K.
Ensslin, Phys. Rev. Lett. \textbf{105}, 116801 (2010).
\end{thebibliography}
\end{document}